\documentclass[12pt]{article}
\usepackage{latexsym}
\usepackage{epsfig}

\usepackage{graphicx}
\usepackage{epstopdf}

\usepackage{epsfig,amssymb,amsmath,amscd}

\newcommand{\bq}{\begin{equation}}
\newcommand{\eq}{\end{equation}}
\newcommand{\bqa}{\begin{eqnarray}}
\newcommand{\eqa}{\end{eqnarray}}
\newcommand{\ben}{\begin{enumerate}}
\newcommand{\een}{\end{enumerate}}
\newcommand{\bc}{\begin{center}}
\newcommand{\ec}{\end{center}}
\newcommand{\bqb}{\begin{eqnarray*}}
\newcommand{\eqb}{\end{eqnarray*}}

\def\pr#1#2#3{Phys. Rev. ${\bf{#1}}$, #2 (#3)}

\def\pl#1#2#3{Phys. Lett. ${\bf{#1}}$, #2 (#3)}

\def\np#1#2#3{Nucl. Phys. ${\bf{#1}}$, #2 (#3)}

\def\epj#1#2#3{Eur. Phys. J. ${\bf{#1}}$, #2 (#3)}

\def\jmp#1#2#3{J. Mod. Phys. ${\bf{#1}}$, #2 (#3)}

\begin{document}
\pagenumbering{arabic}
\thispagestyle{empty}
\def\thefootnote{\fnsymbol{footnote}}
\setcounter{footnote}{1}

\vspace*{2cm}
\begin{flushright}
March. 27, 2018\\
 \end{flushright}
\vspace*{1cm}

\begin{center}
{\Large {\bf Z Polarization in $e^+e^-\to t\bar t Z$
for testing the top quark mass structure and the presence of final interactions.}}\\
 \vspace{1cm}
{\large F.M. Renard}\\
\vspace{0.2cm}
Laboratoire Univers et Particules de Montpellier,
UMR 5299\\
Universit\'{e} de Montpellier, Place Eug\`{e}ne Bataillon CC072\\
 F-34095 Montpellier Cedex 5, France.\\
\end{center}

\vspace*{1.cm}
\begin{center}
{\bf Abstract}
\end{center}

We show that the measurement of the $Z$ polarization
in the  $e^+e^-\to t\bar t Z$ process 
would allow an interesting determination
of the role of the top quark mass.
This can be used for testing the possibility of top compositeness or of
the occurence of final state interactions related to the mass generation 
in particular the interaction with dark matter.

\vspace{0.5cm}

\def\thefootnote{\arabic{footnote}}
\setcounter{footnote}{0}
\clearpage

\section{INTRODUCTION}

We discuss the features of the $Z$ polarization, more especially the 
percentage of $Z_L$ production, in the $e^+e^-\to t\bar t Z$ process.\\
In the standard model (SM), from the equivalence principle \cite{equiv}, 
it is expected that, at high
$p_Z$, the $Z_L$ production rate is equal to the $G^0$ one. The $G^0tt$
coupling being proportional to the top quark mass, a consequence is that the $Z_L$ rate at high $p_Z$ should be proportional to $m^2_t$. We will first review how its precise value depends on the kinematics of the $e^+e^-\to t\bar t Z$ process.\\
But the aim of the present study is to see how this $Z_L$ rate could be
modified by some non standard effects.
A priori it should be related to the way the top quark mass is generated.\\
Top quark (and Higgs boson) compositeness (see discussions in refs.\cite{comp, Hcomp2,Hcomp3,Hcomp4,partialcomp}) could be one example at the origin of a variable
(kinematically dependent) top mass, see \cite{trcomp, CSMrev}. 
We will show how
this would directly reflect in the $Z_L$ rate.\\
Another possibility is the relation between the top quark mass generation
and the presence of dark matter (DM), see \cite{revDM}, \cite{DMmass}.
As mentioned in \cite{DMexch}
this may imply the existence of a special interaction between massive
particles. In our case the presence of a ($Z_Lt$) and a ($Z_L\bar t$)
final state interaction would indeed modify the global $Z_L$ rate.\\
We have presently no precise model for computing this type of effect.
We will only illustrate two phenomenological cases which correspond
to strong effects.\\ 

Contents: In Section 2 we review the SM prediction for the $Z_L$ rate
compared to the $G^0$ rate. In Section 3 we show the effect of 
an effective $m_t(s)$ top quark mass for example due to top quark 
compositeness. In Section 4 we show possible effects of final state 
DM interactions. Concluding remarks and implications for future searches
are  given in Section 5.\\

\section{Z polarization in SM case}

At Born level in SM the $e^+e^-\to t\bar t Z$ process is described by
the 5 diagrams of Fig.1.\\
Each diagram (a) to (e) contributes to both transverse and longitudinal
$Z$ polarization.
A priori the longitudinal polarization is favored by the ${p_Z\over m_Z}$
factor appearing in the $Z$ polarization 4-vector $\epsilon_Z$.
However, for high $p_Z$, there are typical gauge cancellations which occur
within class of diagrams and which avoid high energy divergences. 
In agreement with the equivalence principle, 
the result is equal
(up to ${m^2_Z/p^2_Z}$ terms) to the contribution of similar diagrams 
where the $Z_L$ is replaced by the Goldstone $G^0$ boson.\\
The simplest case appears for the sum of the "$Z$ emission" diagrams (d+e). 
In this sum, for high $p_Z$, the $Z_L$ contribution cancels 
and this is in agreement with the vanishing $G^0ee$ couplings.\\
The other contributions (a+b+c) have mixed properties. Cancellations
appear in the part of (a+b) which does not depend on the top mass; the remaining 
mass dependent (a+b) part
and the (c) contribution, which are proportional to $m_t$, agree with
the contributions of the similar diagrams where $Z_L$ is replaced by $G^0$
(whose $G^0tt$ couplings are proportional to $m_t$).\\
These features can be followed by looking at the $Z_L$ ratio
\bq
R_L={\sigma(t\bar t Z_L)\over \sigma(t\bar t Z_T)+\sigma(t\bar t Z_L)}
\eq
whose properties depend on the relative importance of the $Z$ emission (d+e) 
and of the annihilation (a+b+c) parts.\\

This relative importance depends on the $t\bar t Z$ kinematics.
For example the  $Z$ emission part increases when $\theta_Z$ is close
to zero or $\pi$ due to the behaviour of the electron propagator. 
At high $p_Z$ there is almost exact cancellation
of the $Z_L$ amplitude; but at low $p_Z$, when $m^2_Z$ terms are not negligible,
this cancellation is not complete and $Z_L$ contributions increase.\\
For central $\theta_Z$ values the annihilation (a+b+c) part dominates.
Its $m_t$ independent part leads to $Z_L$ cancellation similarly 
to the above $Z$ emission. But the $m_t$ terms (in agreement
with the equivalence to the contributions of $G^0$ production
diagrams) do not cancel. This finally gives $Z_L$ contributions increasing with $p_Z$
(as one expects from $G^0$ vertices).\\

In the following illustrations we will consider cross sections integrated
over $t$ and $\bar t$ kinematics with cuts avoiding collinear
singularities and we will discuss the behaviour of the corresponding ratio $R_L$.\\

In Fig.2, for two angular values $\theta_Z={\pi\over6},{\pi\over2}$, 
we first compare the behaviour of $R_L$ with the one of the equivalent
ratio
\bq
R_L(G^0)={\sigma(t\bar t G^0)\over \sigma(t\bar t Z_T)+\sigma(t\bar t G^0)}
\eq

We can see, as expected, that
$R_L(G^0)$ coincides with $R_L$ for high $p_Z$.
This is especially true for central $\theta_Z$ values where the
annihilation (a+b+c) part dominates. For collinear values
($\theta_Z\simeq 0$ or $\simeq\pi$) the $Z$ emission (d+e) dominates at low $p_Z$
where the $G^0$ equivalence does not apply because of $m^2_Z$ terms;
consequently large  $Z_L$ fractions appear in this domain.\\

As an additional confirmation about the role of the top quark mass in $R_L$
we have computed this ratio for the $e^+e^-\to b\bar b Z$ process
and drawn the result in Fig.2 for comparison with the $e^+e^-\to t\bar t Z$
case. 
One sees that indeed $R_L(b)$ is much smaller than $R_L(t)$
with the factor $m^2_b/m^2_t$ at high $p_Z$.\\

This review of the SM case shows that the measurement of the $Z$ polarization
constitutes a very powerful test of the top quark mass 
effect in the $e^+e^-\to t\bar t Z$ process. In the next section we consider possible 
new physics effects affecting the top quark mass and reflecting in the
$R_L$ behaviour.

\section{Examples of non standard effects}

\subsection{Effective top quark mass}

A simple example is given by top quark (and possibly Higgs boson)
compositeness. In analogy with the hadronic case its mass may
(at least partly) arise from a binding interaction. In this case
(like in QCD) the effective mass would be scale dependent.\\
This has been mentioned in \cite{trcomp,CSMrev}.\\

We will now show how this possibility would reflect in the
behaviour of the $Z$ polarization rate in the $e^+e^-\to t\bar t Z$ process.\\

In the absence of a precise model,
in our illustrations, for simplicity, we  will first show the consequences of
a unique effective mass $m_t(s)$ where $s$ is the total $e^+e^-$ energy squared

\bq
m_t(s)=m_t{(m^2_{th}+m^2_0)\over (s++m^2_0)}
\eq

In Fig.3 we can see the reduction of $R_L$ generated by the use of
$m_t(s)$ with $m_0=2$ or $4$ TeV.\\

We have also looked at the effect of the use of effective masses
$m_t(x)$ depending on specific subenergies $x=s_{Zt},s_{Z\bar t},s_{t\bar t}$
appearing in each diagram. Similar effects are obtained but obviously
numerically depending on the chosen specific binding scale $m_0$.\\

\subsection{Dark matter final state interaction}

We now follow the assumption that the masses of the heavy SM particles arise from a special interaction with DM environment, \cite{DMmass}; for review about DM see
for example \cite{revDM}.
Consequently one may expect non standard interactions between these
heavy SM particles, \cite{DMexch}. They could be detected as final state interactions
of heavy SM particles produced in standard processes.\\
This can be the case in the present $e^+e^-\to t\bar t Z$ process.
Final state interactions may appear between ($Zt$), ($Z\bar t$) and
($t\bar t$).\\
As discussed in  \cite{DMexch} if such interactions are related to mass generation
they could be specific of the longitudinal gauge bosons (and correspondingly
of the Goldstone bosons).\\ 
In the present process the ratio $R_L$ would be modified by final state processes
$Z_Lt\to Z_Lt$, $Z_L\bar t\to Z_L\bar t$ but not by the $t\bar t\to t\bar t$
one (the identification of this interaction could be done by measurement of the 
top quark polarization in $e^+e^-\to t\bar t$ discussed in \cite{DMexch}).\\
We will now make illustrations first by modifying the $Z_Lt\bar t$
amplitudes by the $(1+C(s_{Zt})) (1+C(s_{Z\bar t}))$ "test factor"
with
\bq
C(x)=1+{m^2_{t}\over m^2_0}~ln{-x\over (m_Z+m_t)^2} ~~, \label{Fs}
\eq
\noindent
$x=s_{Zt}$ or $s_{Z\bar t}$ and $m_0=0.5$ TeV, like in \cite{DMexch}.\\
The result can be seen in Fig.4 with the curves (DMZ) compared to the standard
SM ones.\\
One will also add the possible contribution of the production
$e^+e^-\to t\bar t G^0$ followed by final $G^0t\to Z_Lt$ and 
$G^0\bar t\to Z_L\bar t$ interactions. This increases the effects as shown
by curves (DMZG).\\

These choices are arbitrary but they show that indeed the measurement of
the Z polarization, possibly with variable kinematical conditions, could
reveal interesting mass generation properties.\\

\section{CONCLUSION}

In this paper we have started by reviewing how in SM the $Z$ polarization depends  
on the $e^+e^-\to t\bar t Z$ kinematics and, essentially, how the $Z_L$ proportion is 
proportional to the top quark mass.\\
We have then considered some examples of non standard effects which may affect
this property.
A first example is the case of top quark compositeness  where an effective 
scale dependent top quark mass would depend on the kinematics and produce
a change of the $Z_L$ proportion.
We have given illustrations with different values of the binding scale
indeed showing immediate modifications of the SM prediction.\\
As a second example we assume that heavy particles (in our case the $Z$,
$t$, $\bar t$), whose masses would be generated by the DM environment, get
final state interactions after their standard production. Illustrations
with some arbitrary choices of parameters show that different modifications of the
$Z_L$ proportion could appear with specific kinematical dependences.\\
Experimental possibilities should be studied for future $e^+e^-$
colliders with precise choices of detection characteristics, collinear
cuts, radiative corrections,... . The present physics expectations at high energy
$e^+e^-$ colliders can for example be found in \cite{Moortgat}.\\
Further developments of our study for other colliders are in progress.\\

\newpage

\begin{figure}[p]
\vspace{-0cm}
\[
\hspace{-3cm}\epsfig{file=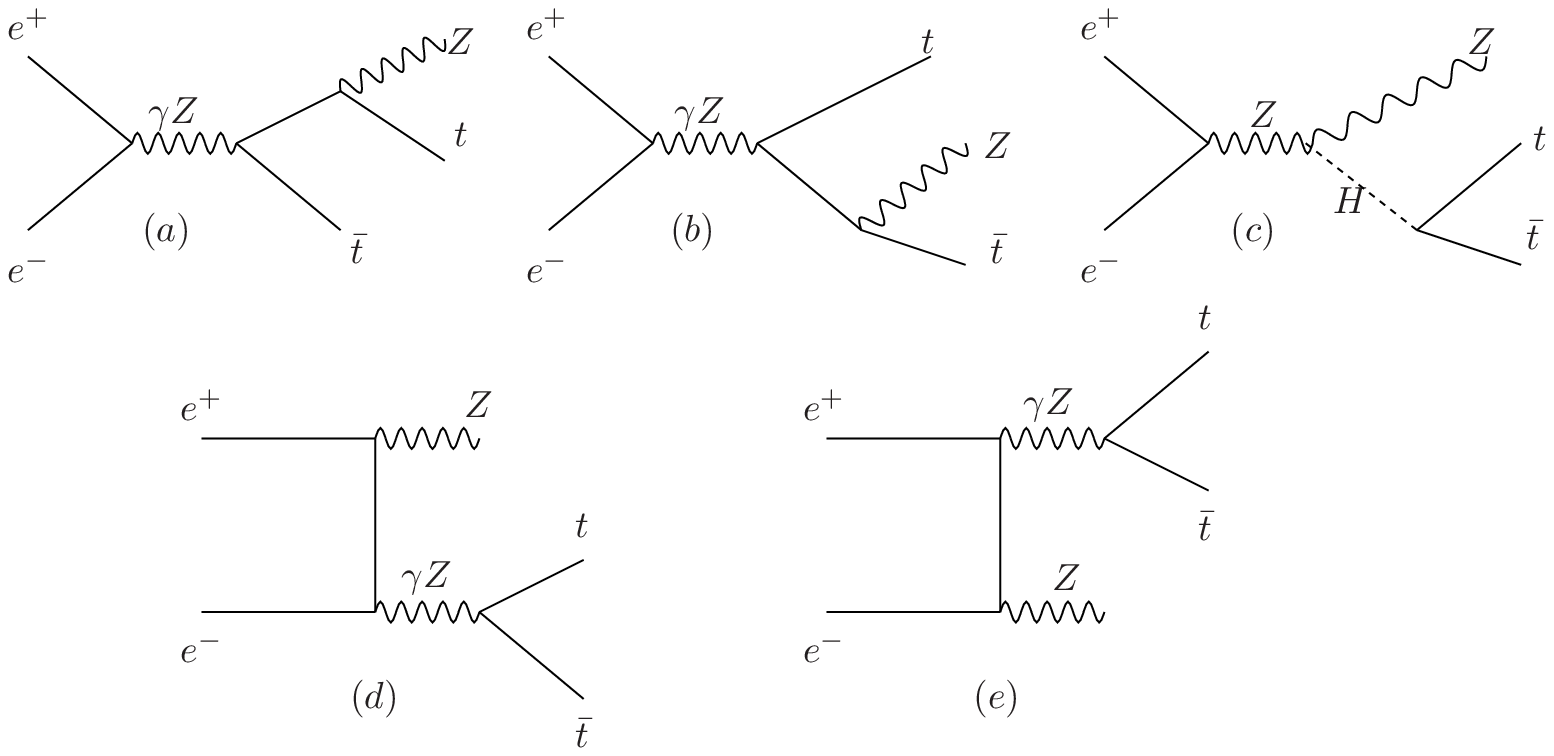 , height=30.cm}
\]\\
\vspace{-17cm}
\caption[1] {SM Born diagrams for $e^+e^-\to t\bar t Z$.}
\end{figure}
\clearpage

\begin{figure}[p]
\vspace{0cm}
\[
\hspace{-2cm}\epsfig{file=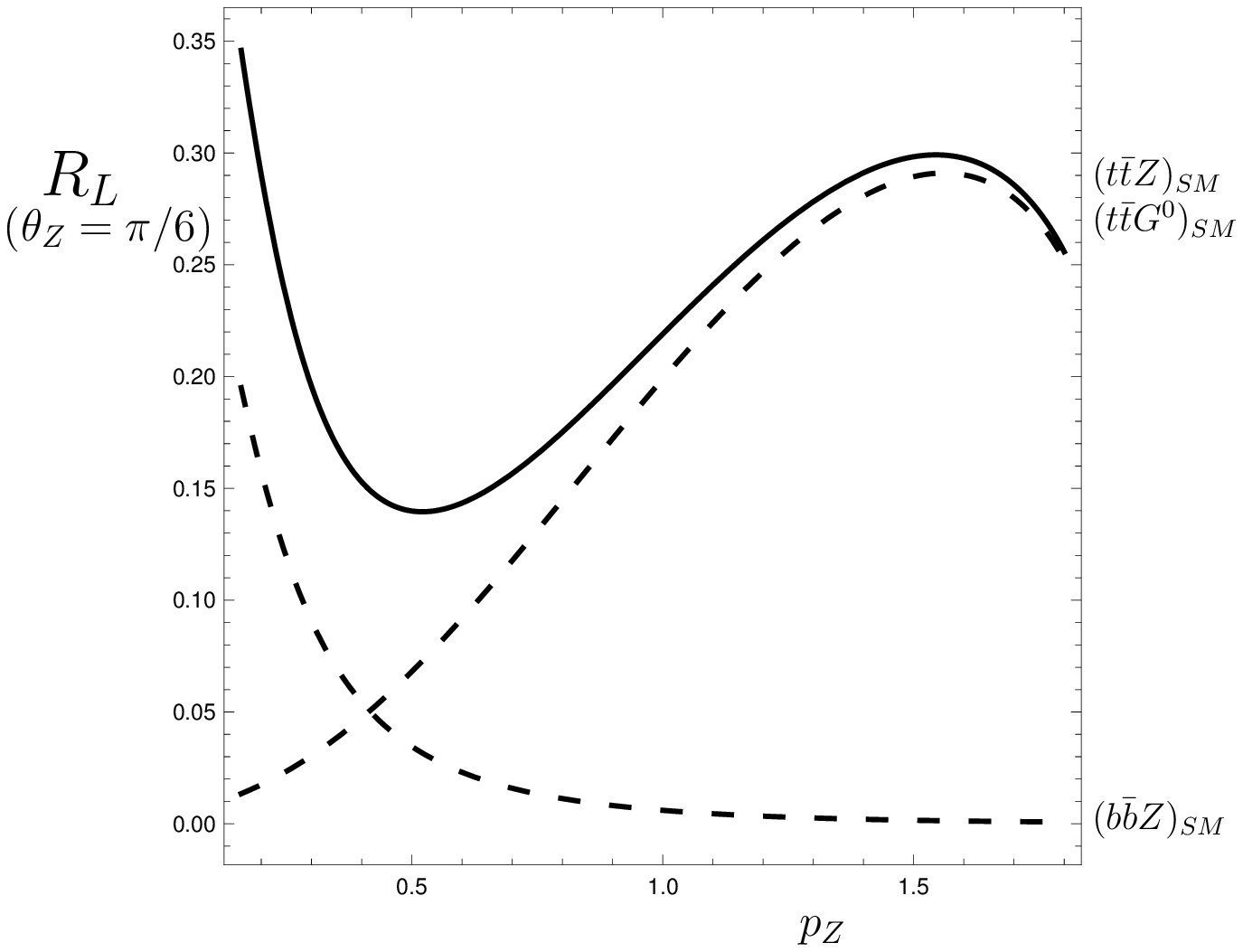 , height=20.cm}
\]\\
\vspace{-13cm}
\[
\hspace{-2cm}\epsfig{file=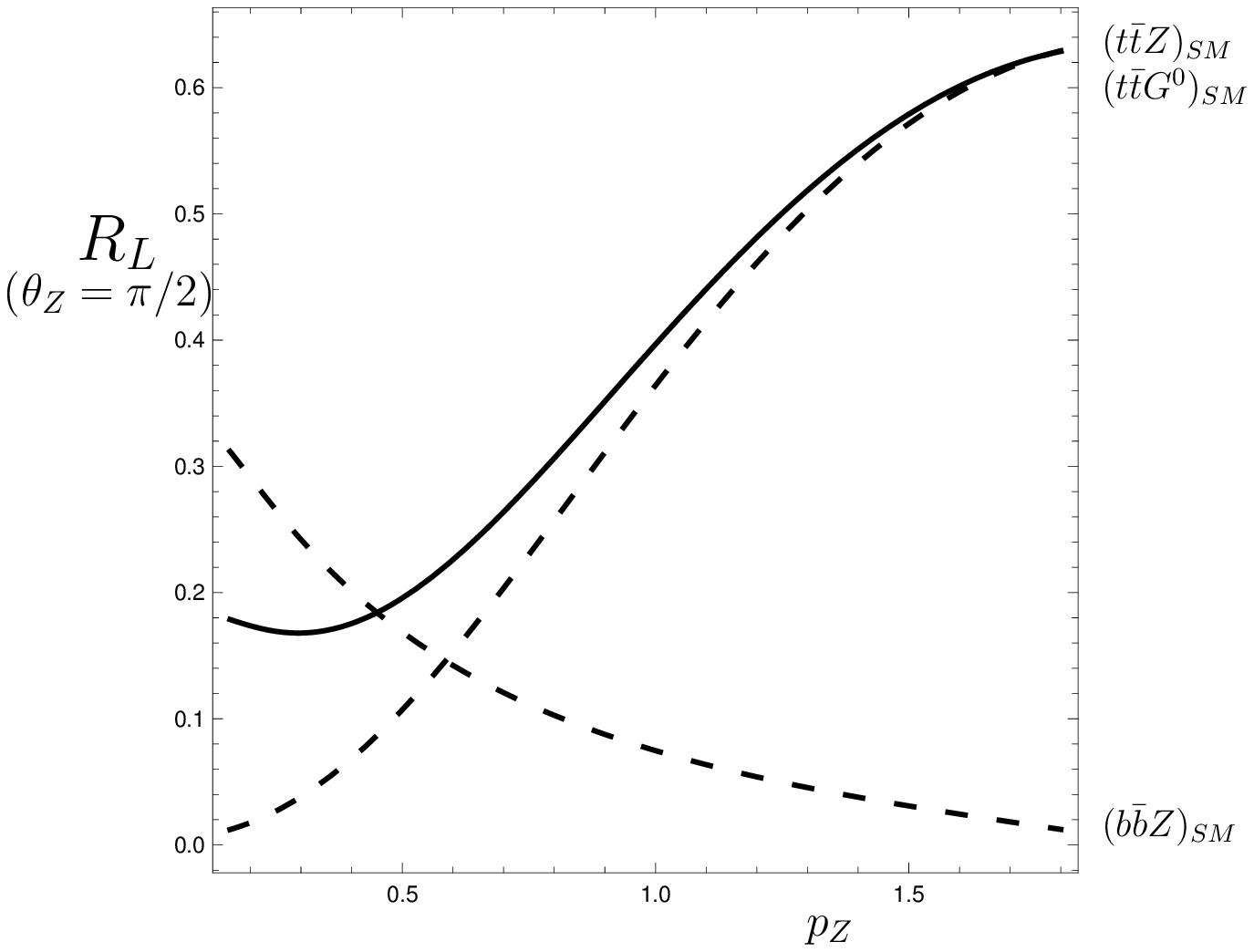 , height=20.cm}
\]\\
\vspace{-10cm}
\caption[1] {$Z_L$ ratio in $e^+e^-\to t\bar t Z$ compared to $G^0$ ratio and
to the $e^+e^-\to b\bar b Z$ case.}
\end{figure}

\clearpage

\begin{figure}[p]
\vspace{-0cm}
\[
\hspace{-2cm}\epsfig{file=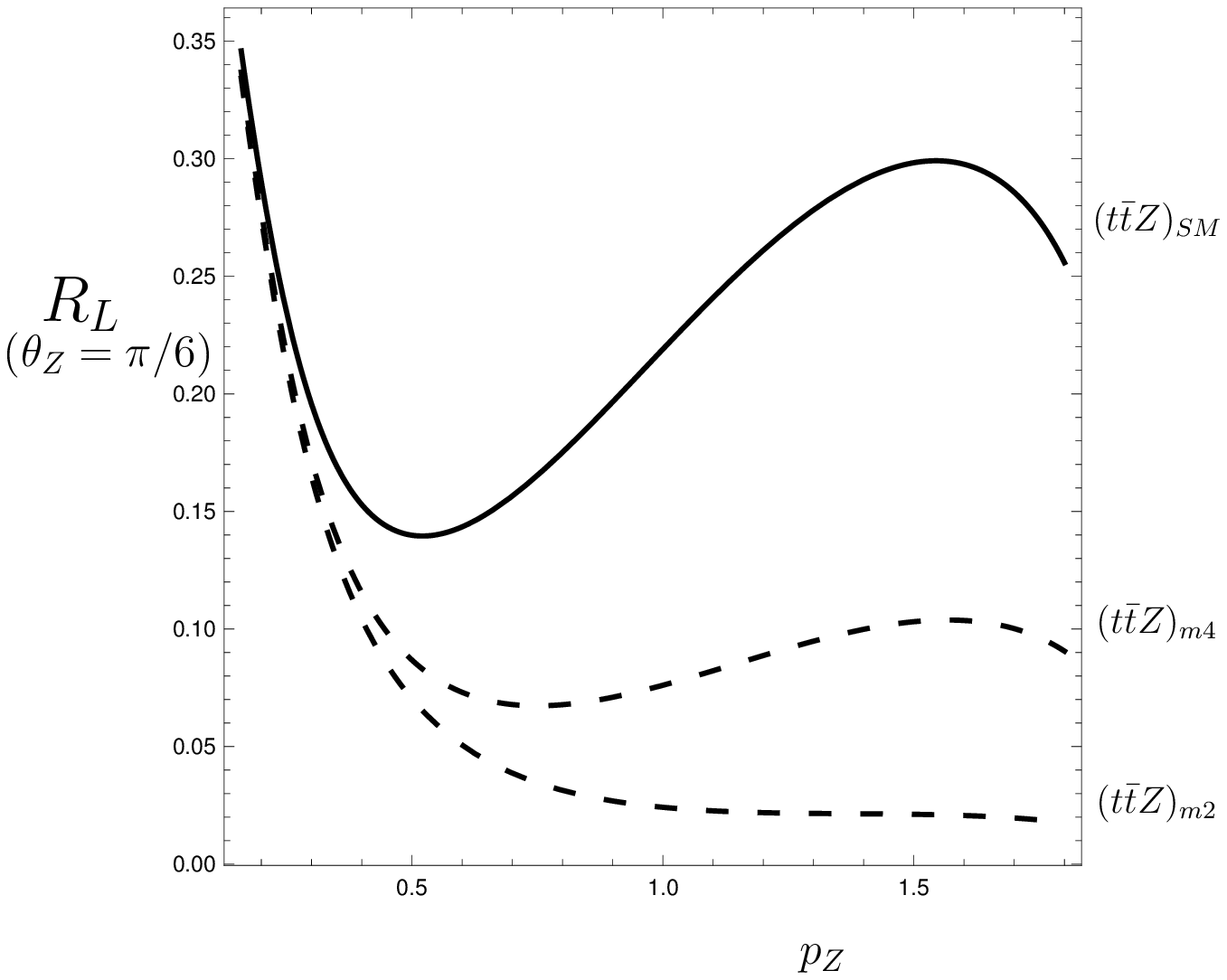 , height=20.cm}
\]\\
\vspace{-13cm}
\[
\hspace{-2cm}\epsfig{file=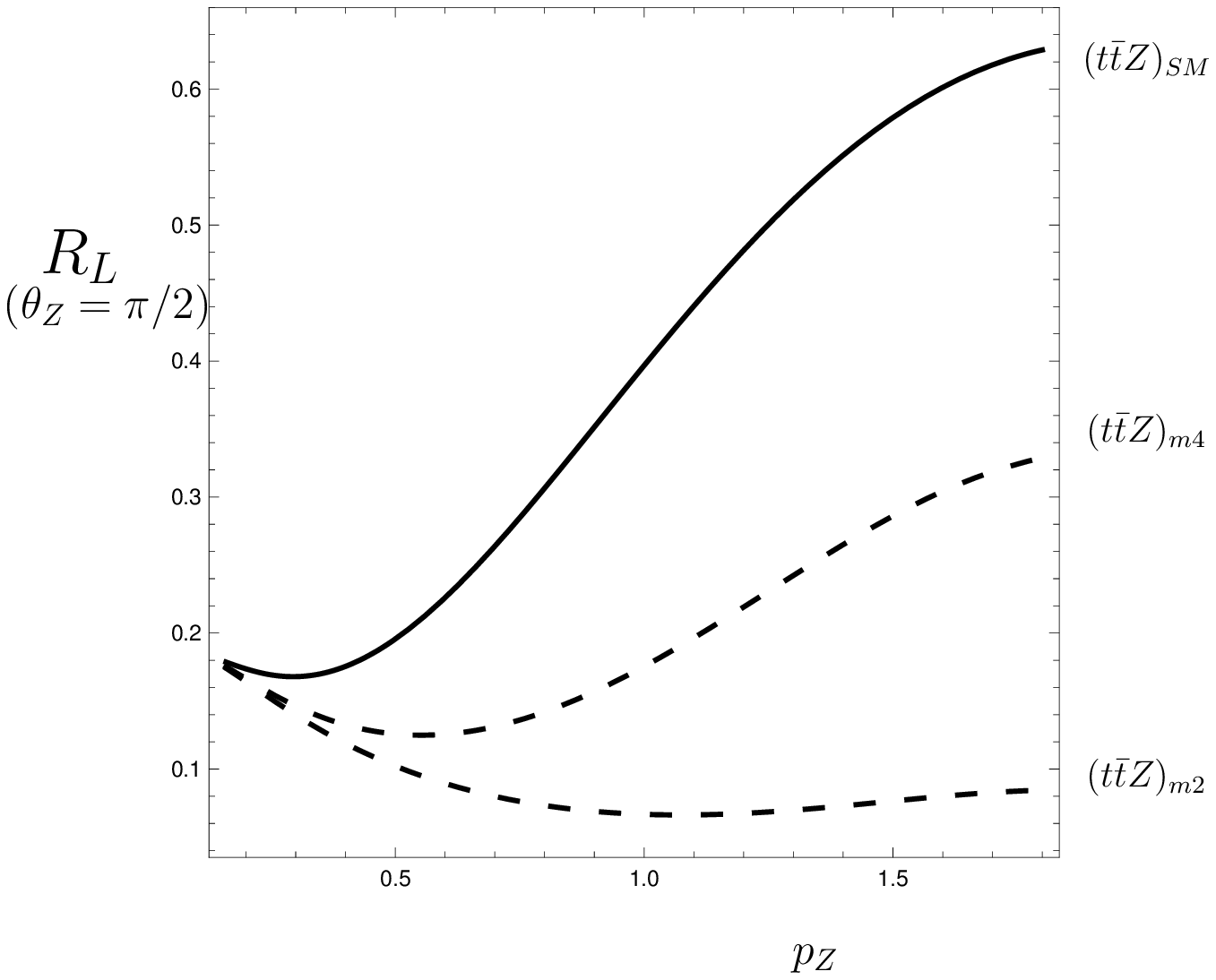 , height=20.cm}
\]\\
\vspace{-10cm}
\caption[1] {$Z_L$ ratio in $e^+e^-\to t\bar t Z$ for 3 cases of $m_t$: SM value and effective
values with a scale at 2 and at 4 TeV.}
\end{figure}
\clearpage

\begin{figure}[p]
\vspace{-0cm}
\[
\hspace{-2cm}\epsfig{file=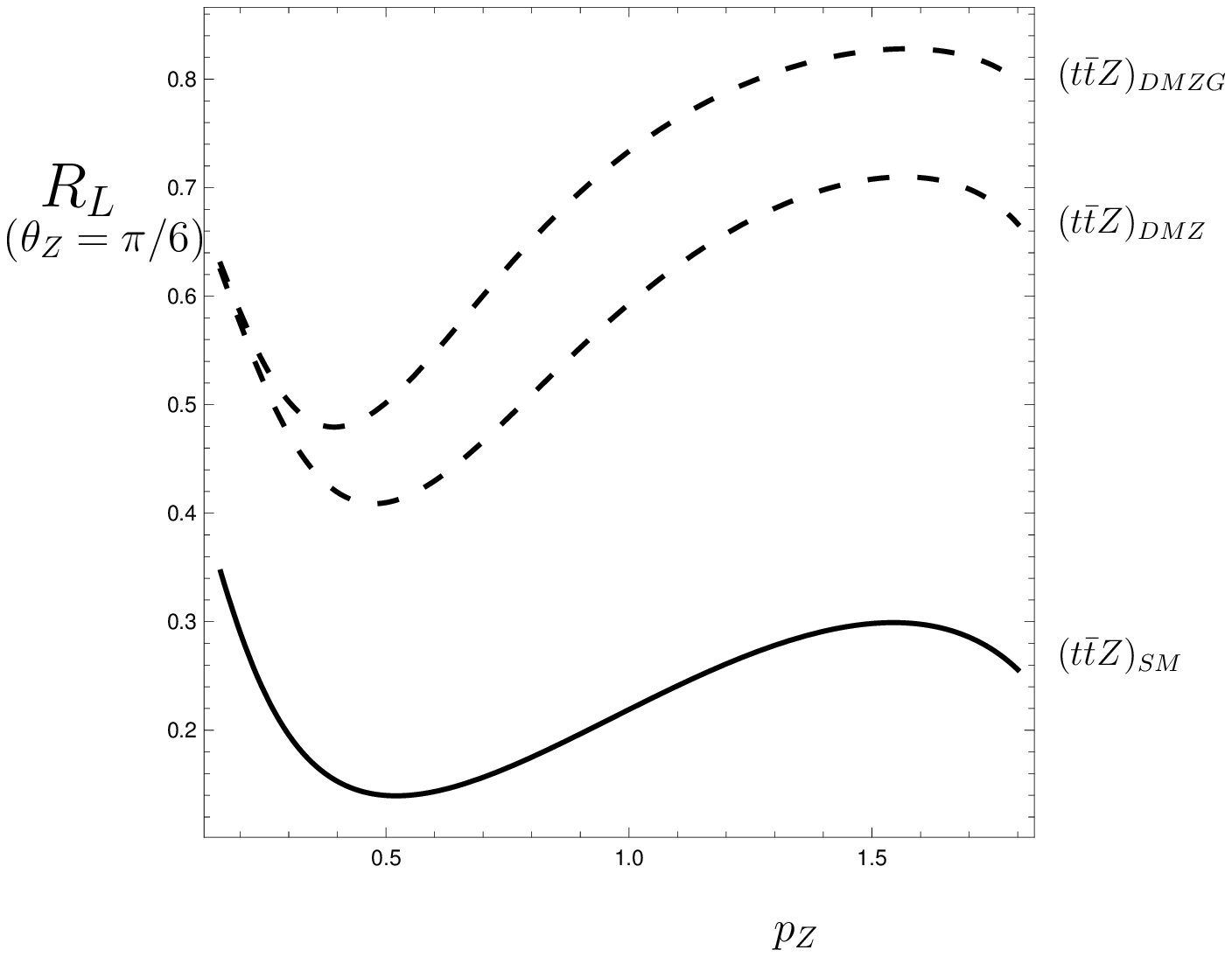 , height=20.cm}
\]\\
\vspace{-13cm}
\[
\hspace{-2cm}\epsfig{file=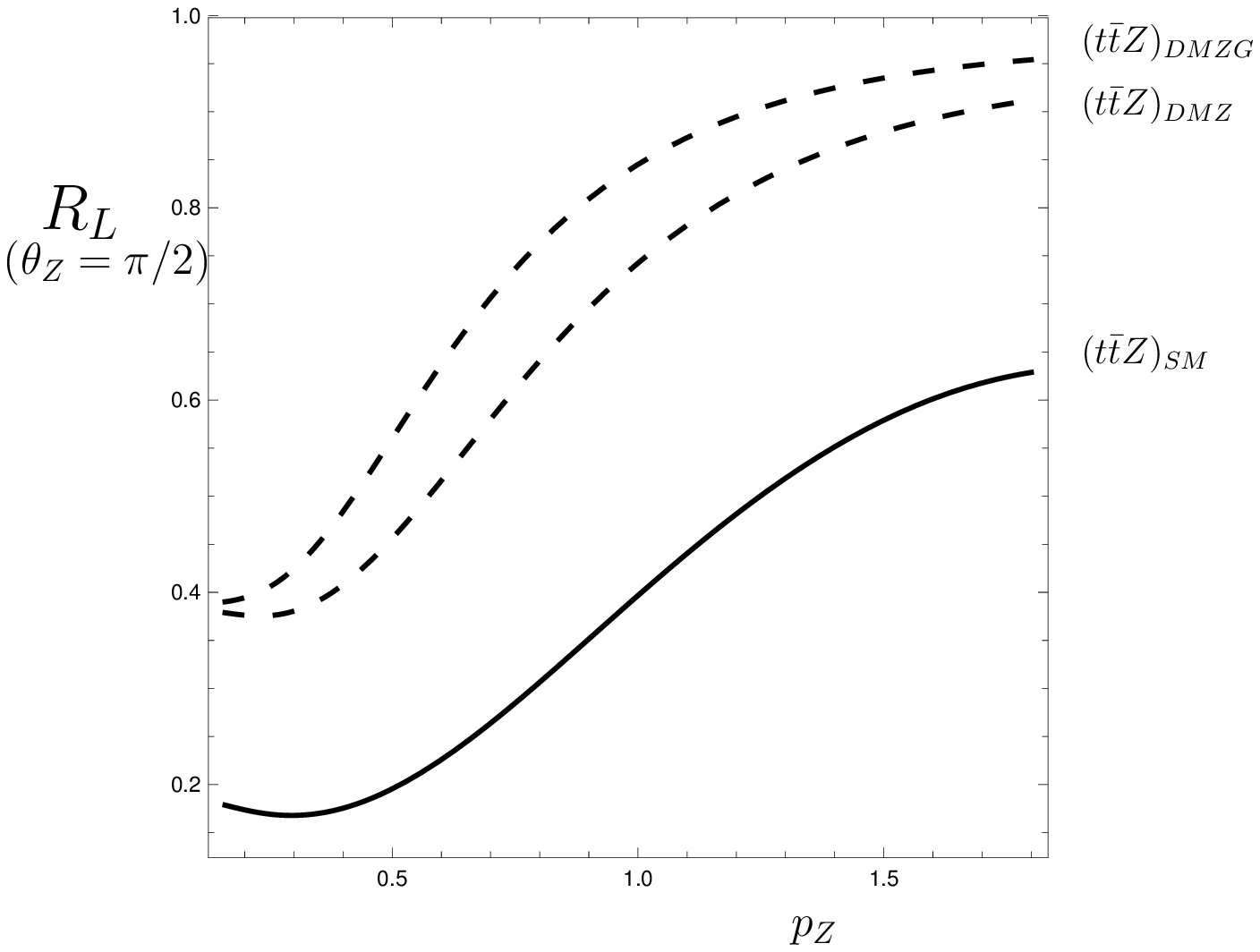 , height=20.cm}
\]\\
\vspace{-10cm}
\caption[1] {$Z_L$ ratio in $e^+e^-\to t\bar t Z$ with final DM $Zt(\bar t)\to Zt(\bar t)$ interaction 
and in adition with final $G^0t(\bar t)\to Zt(\bar t)$ interaction.}
\end{figure}
\clearpage

\clearpage


\begin{thebibliography}{99}
%
\bibitem{equiv}
J.M.Cornwall, D.N.Levim and G.Tiktopoulos, Phys. Rev.D10(1974)1145 ;
D11(1975) 972E; C.E.Vayonakis, Lett. Nuovo Cimento17(1976) 383;
B.W.Lee, C.Quigg and H.Thacker, Phys. Rev.D16(1977) 1519 ;
M.S.Chanowitz and M.K.Gaillard, Nucl. Phys.B261(1985) 379;
M.S.Chanowitz, Ann.Rev.Nucl.Part.Sci.38(1988)323;
G.J.Gounaris, R.Koegerler and H.Neufeld, Phys. Rev.D34(1986) 3257.
%
\bibitem{comp}  H. Terazawa, Y. Chikashige and K. Akama, \pr{D15}{480}{1977};
for other references see
H. Terazawa and M. Yasue, Nonlin.Phenom.Complex Syst. {\bf19},1(2016);
\jmp{5}{205}{2014}.
%
\bibitem{Hcomp2} D.B. Kaplan and H. Georgi, \pl{136B}{183}{1984}.
%
\bibitem{Hcomp3} K. Agashe, R. Contino and A. Pomarol, \np{B719}{165}{2005}; hep/ph 0412089.
%
\bibitem{Hcomp4} G. Panico and A. Wulzer, Lect.Notes Phys. {\bf 913},1(2016).
%
\bibitem{partialcomp} R. Contino, T. Kramer, M. Son and R. Sundrum,
J. High Energy Physics {\bf 05}(2007)074.
%
\bibitem{trcomp}  G.J. Gounaris and F.M. Renard,
arXiv: 1611.02426.
%
\bibitem{CSMrev}  F.M. Renard, arXiv: 1708.01111.
%
\bibitem{revDM} B. Penning, arXiv: 1712.01391. We also thank Mike Cavedon
for interesting informations about this subject.  
%
\bibitem{DMmass} F.M. Renard, arXiv: 1712.05352.
%
\bibitem{DMexch} F.M. Renard, arXiv: 1801.10369. 
%
\bibitem{Moortgat} G.  Moortgat-Pick et al, \epj{C75}{371}{2015}, arXiv: 1504.01726.
%



\end{thebibliography}
\end{document}